# Quantum tomography of free electrons

Y. Fang[1,2,†], J. Kuttruff[3,†], Z. Zhao[2], L. Moehrle[3], P. Hommelhoff[1,2,*], P. Baum[3,*]

[1]*Faculty of Physics, Ludwig-Maximilians-Universität München, Munich, Germany*
[2]*Department of Physics, Friedrich-Alexander-Universität Erlangen-Nürnberg (FAU), Erlangen, Germany*
[3]*Universität Konstanz, Fachbereich Physik, 78464 Konstanz, Germany*
[†]*These authors contributed equally to this work*
[*]*peter.hommelhoff@fau.de*
[*]*peter.baum@uni-konstanz.de*

**Determining the quantum state of a given quantum-mechanical system is a fundamental task in physics. Quantum-state tomography has been pivotal for establishing quantum optics [1-4] and for revealing the properties of bound charges in materials [5-7]. An emerging other object for studying and utilizing quantum effects are free electrons, elementary particles that are central to high-resolution microscopy [8,9], electron-based quantum optics [10-17], ultrafast electron microscopy [18-24] and particle accelerators [25-27]. However, free electrons are intrinsically incoherent, and we lack a broadly applicable method to measure and control their quantum state beyond special cases with discrete energy sidebands [28,29]. Here, we report a universal approach to measure arbitrary free-electron quantum states in continuous variables. Two monochromatic but spectrally shifted laser waves produce interfering quantum paths that directly reveal the density matrix and thus all essential properties of the pure wavepackets, the ensemble, and their interlinks. As a first application, we show how the quantum state of a single electron is modified by many-body Coulomb interactions of a surrounding electron gas. The reported concepts and results provide insight into otherwise hidden correlations in electron beams and enable the controlled optimization of exceptional quantum states for free-electron quantum optics or quantum electron microscopy.**

Figure 1a shows the concept of our experiment. Electron pulses are created by photoemission from a Schottky field-emitter source in a transmission electron microscope [30,31], operated at an electron central kinetic energy of $E_{el}$=180 keV. Such electron sources are widely used in almost all ultrafast electron microscopy [18-24] and other numerous free-electron applications [10-17,26-33]. The emerging femtosecond electron pulses are suspected to be in a mixed state composed of pure electron wavepackets with statistically distributed arrival times and central energies [34]. The degree of coherence in energy and time is determined by the physics of the emission process, voltage fluctuation of the instrument and statistical Coulomb interactions with other electrons [28]. Due to these processes, the phase space volume of the free electrons is expected to be larger than that of the underlying pure electron states (top left inset). Chirp and other correlations between the ensemble and the individual coherent parts as functions of time and energy are unknown. In such a case, the most general quantity to describe the electron's quantum state is the density matrix.

To measure the density matrix, we superimpose the electrons under scrutiny with two phase-stabilized probe laser pulses with ultra-narrow linewidth and slightly detuned wavelengths $\lambda_1$ and $\lambda_2$ (bottom left inset) and use an electron analyzer to measure the electron spectrum as a function of the spectral separation between the two laser waves and the time delay $\Delta t$ between the electron pulses and the combined laser field. We thus sample the density matrix by spectral shearing interferometry in an adjustable basis set.

Figure 1b depicts the principle of our measurement. For each pure electron state $\psi$ in continuous energy variables, interaction with one of the two quasi-monochromatic laser pulses leads to the absorption or emission of a photon, coherently displacing the energy spectrum (filled shape) and the spectral phase (floating curves). The simultaneous interaction of an electronic state with two slightly detuned laser fields of wavelengths $\lambda_1$ and $\lambda_2$ then leads to two final states (solid and dashed) separated by $\Delta E_{ph}=hc/\lambda_1-hc/\lambda_2$. The shifted wave packet $\psi(\varepsilon)$ at ±1 photon energy coherently interferes with its duplicate $\psi(\varepsilon')$ at a slightly different shift, giving rise to an interference pattern in the form of $S(\varepsilon,\Delta t,\Delta E_{ph})=|\psi(\varepsilon)|^2+|\psi(\varepsilon')|^2+2|\psi(\varepsilon)||\psi(\varepsilon')|\cdot\cos[\Delta\phi(\varepsilon)+\Delta E_{ph}\Delta t/\hbar]$, where $\varepsilon$ and $\varepsilon'$ are electron energies with $\varepsilon'-\varepsilon=\Delta E_{ph}$ and where $\Delta\phi(\varepsilon)$ is the phase difference between $\psi(\varepsilon)$ and $\psi(\varepsilon')$. By scanning $\Delta t$, the relative phase between the two shifted wave packets is varied, leading to a sinusoidal intensity modulation whose modulation phase with $\Delta t$ directly reflects the spectral phase (floating curves) as a function of energy. Together with an independent measurement of the spectral amplitude, we thus obtain a full reconstruction of the unknown pure state $\psi$, in analogy to the spectral shearing protocols for optics [35] or pure electron states [36]. For the more general case of mixed states, we obtain similarly to the above expression the signal strength (Methods)

$$S(\varepsilon,\Delta t,\Delta E_{\mathrm{ph}}) \propto \rho(\varepsilon,\varepsilon)+\rho(\varepsilon',\varepsilon')+2\,|\,\rho(\varepsilon,\varepsilon')\,|\cos\left\{\arg[\rho(\varepsilon,\varepsilon')]+\Delta E_{\mathrm{ph}}\Delta t/\hbar\right\}, \qquad (1)$$

now with the density matrix $\rho(\varepsilon,\varepsilon')$. We see that a measured interferogram under varying $\varepsilon$ provides direct insight into the full density matrix: The contrast of the measured interference fringes determines $|\rho(\varepsilon,\varepsilon')|$, and the phase of the measured interference fringes determines $\arg[\rho(\varepsilon,\varepsilon')]$. In practice, the complete density matrix can be obtained by systematically scanning the easily adjustable frequency difference $\Delta E_{\mathrm{ph}}$ and the time delay $\Delta t$.

Figure 1c shows a first proof-of-principle experiment. Monochromatic laser light is created with Fabry-Perot filters (Methods) at $\lambda_1$=514 nm and $\lambda_2$=515.9 nm, corresponding to $\Delta E_{\mathrm{ph}}$=8.9 meV. Using an electron-transparent silicon nitride membrane to mediate the light-electron interaction [21], we produce coherent electron energy sidebands (see Fig. 1c). After a peak at $\Delta t \approx 500$ fs, the amplitude decays slowly with $\Delta t$, reflecting the exponential temporal response of the Fabry-Perot filters (Methods). The measured periodic modulation of the sideband intensity with $\Delta t$ directly evidences the two-path interference between electrons shifted by $\varepsilon$ and $\varepsilon+\Delta E_{\mathrm{ph}}$. A central result of the experiment, reproduced by theory (Fig. 1d), is the strong tilt of the energy sidebands within each temporal interference lobe. Even without full reconstruction (see below), this tilt gives some direct insight into the phase of the density matrix of our electrons. According to Fig. 1b, a spectral phase with constant or linear terms would either interfere constructively or destructively for all spectral components when scanning time. Only a slowly varying spectral phase with nonlinearities (as depicted in Fig.1b) can produce a monotonic shift of the point of constructive interference when scanning time. The measured linear tilt of the features therefore implies a quadratic spectral phase of the electrons (Fig. 1b), corresponding to a linear chirp in the density matrix, as it can be expected from the dispersion of vacuum and space charge effects [34].

Full density matrix reconstruction requires a continuous basis change that we achieve by a tomographic scan over $\Delta E_{\mathrm{ph}}$ and $\Delta t$ (Extended Data Figure 1). Figure 2 shows such a series of measurements for detunings $\Delta E_{\mathrm{ph}}$ between 2 meV and 12 meV. For each electron interferogram, we record a corresponding optical spectrum (Extended Data Figure 2) to determine $\Delta E_{\mathrm{ph}}$. All measured results have a similar rapid rise and slow decay of general signal strength, given by the temporal pulse shape of the optical fields. However, the measured interference shows a transition from a slow modulation for a small detuning to a more rapid oscillation for a larger detuning (Fig. 2a), directly reflecting the oscillating term $\Delta E_{\mathrm{ph}}\Delta t/\hbar$ in Eq. (1). The interference contrast is close to unity for small detunings but almost vanishes for the largest detunings (Fig. 2a), indicating the onset of incoherent effects for electron energy components that differ by a larger $\Delta E_{\mathrm{ph}}$. For a very small detuning, the photon-order peak tilt of the sidebands is weak, but it becomes more and more visible as the detuning grows (Fig. 2a), because the phase differences between $\varepsilon$ and $\varepsilon+\Delta E_{\mathrm{ph}}$ become larger with an increasing $\Delta E_{\mathrm{ph}}$ (see Eq. 1).

We now show how such a systematic data set allows us to reconstruct the complete density matrix in the time-energy domain. The diagonal elements for $\varepsilon=\varepsilon'$ are obtained from an electron spectrum without optical modulation, for example the measured spectrum at $\Delta t=0$ in Fig. 2a. The amplitudes of the off-diagonal elements are obtained by extracting the measured oscillation amplitude, that is, the contrast of the time-domain interference below the rapidly increasing and slowly decreasing envelope from the Fabry-Perot filters (Extended Data Figure 3). The phase of the density matrix is reconstructed by extracting the relative phase of the time-domain interference as a function of energy (Extended Data Figure 3). Here, we align the phase of the matrix elements at $\varepsilon'+\varepsilon=0$ to zero to compensate for the slight systematic phase offset among different measurements.

Figure 2b shows the reconstructed amplitude of the density matrix, and Fig. 2c shows the reconstructed phase. The remaining white areas in both Figures correspond to elements that are not directly covered by the discrete set of measurements. Nevertheless, we already see the general structure of the density matrix of our electrons. The amplitude is elliptical, has no complicated inner structure, and lies horizontally along the direction of $\varepsilon'+\varepsilon$ axes. The measured elliptical aspect ratio between the $\varepsilon'-\varepsilon$ and $\varepsilon'+\varepsilon$ axes is $\eta\approx8\times10^{-3}$, which is substantially smaller than for a pure state, where $\eta=1$ [7]. The extension along the $\varepsilon'-\varepsilon$ axes reveals the degree of quantum coherence of our electrons, which is ~14 meV (full width at half maximum). This value is ~60 times smaller than the direct readout of the spectral width with the electron energy analyzer, ~0.9 eV (see $\Delta t=0$ in Fig. 2a). The reconstructed phase shows opposite dependencies on $\varepsilon'+\varepsilon$ on the two sides of $\varepsilon'-\varepsilon=0$; the farther the elements are from $\varepsilon'-\varepsilon=0$, the more rapidly the phase varies as a function of $\varepsilon'+\varepsilon$. Throughout the entire density matrix, the phase is continuous and does not have oscillations or discontinuities. The electrons in our experiment have therefore smooth but non-trivial correlations between their coherent and incoherent widths in energy and time.

Next, we close the gaps in the reconstructed density matrix. We describe the emitted electron pulses by four key parameters (Methods): We consider the pure energy bandwidth, which is linked to the temporal coherence length, an incoherent jittering in time, an incoherent jittering in energy, and a common linear chirp for pure wavepacket and ensemble. Some of these quantities are linked to the easily measurable aspect ratio $\eta$ of the density matrix (Methods). To facilitate the convergence of the multidimensional fit, we fix the time jitter to 277 fs [37] and limit $\eta=8\times10^{-3}\pm20\%$ to avoid unrealistic local minima (Methods). Fitting with a maximum likelihood estimation to the measured density matrix elements (Figs. 2b-c) fills the voids in the measurement (Figs. 2d-e) and reveals a pure energy bandwidth of 0.13±0.04 eV, a temporal coherence length of 14±5 fs, an incoherent energy spread of 0.88±0.03 eV, a common chirp of -94±2 fs/eV (Fig. 2f), and an aspect ratio $\eta$=(7.8±0.2)×$10^{-3}$. We see that the measured electrons are in a partially mixed state: the temporal coherence length is about twenty times shorter than the electron pulse duration and the pure energy spectrum is about ten times narrower than the incoherent bandwidth of the beam. These results provide insight into the physics of electron emission at our emitter tip. Individual emitted electrons are quantum-limited wave packets with a measured pure bandwidth (130 meV) and coherence length (14 fs) matching the expected transition bandwidth (140 meV) and uncertainty-limited emission time (13 fs) of the relevant part of the band structure [31]. Such individual electrons are randomly emitted in time within ~277 fs, and their central energy jitters by ~0.88 eV due to an inhomogeneous Schottky effect, fluctuations of the work function by surface effects, or space charge forces inside or outside of the emitter material. Propagation then produces chirp in a similar way for the individual wavepackets and for the ensemble. This means that the typical separation between single electrons is larger than their coherence length, such that electrons do not substantially overlap as two-electron states [28].

According to theory, this spectral shearing interferometry is suitable for quantum state tomography of almost any free-electron quantum states. For example, we should be able to reconstruct an electron cat state of the form $|\text{cat}\rangle \propto |E_{\text{el}}+\Delta E\rangle + |E_{\text{el}}-\Delta E\rangle$, where $E_{\text{el}}$=180 keV and $\Delta E$=0.3 eV. Such a state could be useful in quantum metrology [12,13] and can be generated if a single-electron emitter is driven by a coherent two-color field. Figure 3a shows the density matrix of such a cat state and Fig. 3b its Wigner distribution [38]. To demonstrate that our method can reconstruct this cat state, we simulate the expected experimental results (Figs. 3c and 3d and Extended Data Fig. 4) with realistic spectral resolution and noise. We see that a small detuning $\Delta E_{\text{ph}}$=70 meV probes the coherence *within* each $|E_{\text{el}}+\Delta E\rangle$ and $|E_{\text{el}}-\Delta E\rangle$ (Fig. 3c), while a larger detuning $\Delta E_{\text{ph}}$=620 meV probes the coherence *between* $|E_{\text{el}}+\Delta E\rangle$ and $|E_{\text{el}}-\Delta E\rangle$ (Fig. 3d). Consequently, we obtain insight into the density matrix by following the above-described procedure and can thus fully reconstruct the density matrix (Fig. 3e) reproducing the initial quantum state almost perfectly (Fig. 3a). Slight artifacts are due to multiphoton-effects in the electron-light interaction, which could be avoided by using weaker laser fields. The reconstructed Wigner distribution (Fig. 3f) reproduces the initial quantum state and its negative values very well.

In ultrafast electron microscopy and quantum optics, single electrons with high temporal coherence are needed for attosecond time resolution [21-23], free-electron quantum walks [10-17] or quantum entanglement [16,17,14,28], but temporal coherence might be reduced by space charge effects in realistic electron beams. As a first application of our free-electron spectral shearing interferometry, we thus measure how an increasing amount of many-body Coulomb interactions at a nanometer-sized emitter tip alters the quantum state and density matrix of a single, post-selected free electron.

Figure 4a shows the resulting raw data and Fig. 4b shows zooms into one interference lobe (red squares in Fig. 4a). We apply a fixed $\Delta E_{\mathrm{ph}}$ of 6.5 meV. We directly see that the tilt of the measured interference features (solid line) increases with electron density. Using similar methods as above, we obtain insight into the density matrix. At $\varepsilon'-\varepsilon$=6.5 meV, the extracted phases $\Phi(\varepsilon)=\arg[\rho(\varepsilon,\varepsilon+\Delta E_{\mathrm{ph}})]$ as a function of electron density (Fig. 4c) are mostly linear with energy but their slopes increase from ~0.9 rad/eV for approximately one electron in the cloud (green) to ~1.1 rad/eV for ~15 electrons (purple), to ~1.3 rad/eV for ~40 electrons (blue) and to ~2 rad/eV for ~150 electrons (red). For further insight, we simulate the density matrix of the electrons under scrutiny with a semiclassical model (see Methods and Extended Data Fig. 5). We approximate the space-charge field as a continuous charge density [39] and simulate the quantum-mechanical propagation of $10^4$ individual pure-state electrons at different emission times and emission energies by solving the Hamilton-Jacobi equation and Feynman's path integral [40,41]. Energetic and temporal widths are taken from our measured results (Fig. 2f); the initial spectral phase of the pure states is assumed to be zero [31]. Finally, we obtain the density matrix of the ensemble by averaging over the density matrices of all pure states. Afterwards, we extract simulated experimental observables (Methods). Figure 4d shows the resulting density matrix phase $\Phi$ at $\Delta E_{\mathrm{ph}}$=6.5 meV in comparison to the measurement (Fig. 4c).

Figures 4e-f show a matrix of the simulated pure states (colored ellipses) together with the shape of the incoherent mixture (grey) in absence of the Coulomb interaction (Fig. 4e) and for 150 electrons per pulse (Fig. 4f). Without Coulomb forces (Fig. 4e), all pure wavepackets have almost the same phase-space shape (Extended Data Fig. 6), almost independently of injection time and energy. We observe that the chirp of the pure states (tilt of colorful ellipses) is similar to the chirp of the incoherent electron ensemble (white dashed line). We argue that both arise from the same physics, namely dispersion of vacuum during acceleration and propagation in free space [34] that affects a single coherent electron and its jittering temporal and energetic properties in almost the same way [34,39]. A slight nonlinearity at low energies (left part of Fig. 4e) is caused by nonlinear dispersion at very low energy in the first few femtoseconds after emission from the tip. With space charge (Fig. 4f), electrons that are released earlier (bottom rows) or later (top rows) from the tip are accelerated or decelerated by Coulomb effects, creating a substantially enlarged ensemble chirp

(white dashed line), appearing in the experiment as the increased phase slopes of the density matrix with electron density (Fig. 4c). Interestingly, the microscopic chirp of the pure electron wavepackets (tilt of the colorful ellipses) is only slightly affected, and pure energy bandwidth and temporal coherence length remain almost the same (Extended Data Fig. 6). Although earlier and faster electrons are substantially accelerated with respect to slower and less energetic electrons in an incoherent way, the coherent wavepacket properties of each single electron remain almost unaffected by this effect. Experiments that rely on temporal coherence, for example attosecond electron microscopy [21-23], single-electron [10,20] and two-electron quantum walks [28], free-electron qubits [12,13] or coherently shaped matter waves [24] are therefore surprisingly robust to a moderate amount of space charge effects. In contrast, applications that rely on average electron density, for example ultrafast electron diffraction [33], laser-driven particle accelerators [25-27], or terahertz compression [19], are strongly affected by such Coulomb effects.

The combined results show that spectral shearing interferometry with variable basis change can reveal the complete density matrix of free electron in the time-energy domain and thus provide the necessary means to understand and generate useful free-electron quantum states. We can reconstruct almost any quantum state that has no correlations and discontinuities closer and faster than the spectral and temporal resolution of the experiment, given by the spectral separation $\Delta E_{\mathrm{ph}}$ and the cycle period of the applied light waves. The method can be easily extended to other degrees of freedom, for example transverse dimensions, by utilizing the quantized deflection of laser-modulated electrons in slanted geometries [42,43]. We can also measure continuous electron beams with or without time-energy modulation [44] by using continuous probe waves. The demonstrated technique can be readily generalized to multi-electron quantum states [28] or two-path sources [45], providing diagnostics for potentially entangled electrons [28]. The reported connection between the quantum coherence of free electrons and their jittering provides a clear route toward generating free-electron pulses in a nearly pure state that will advance ghost imaging [46,47] or qubit protocols [12,13]. Necessary basis for any such endeavor in the fast-growing research field of free-electron quantum optics is a universally applicable diagnostic technique, now available.

**Acknowledgements:** We acknowledge funding by Deutsche Forschungsgemeinschaft (DFG) via SFB 1432 and projects HO 4543/5 & HO 4543/8, by the European Research Council (ERC) via

AdG ULMI, and by the Gordon and Betty Moore Foundation via GBMF11473. Z. Z. acknowledges the support from the Alexander von Humboldt Foundation (Postdoctoral Fellowship).

**Author contributions:** Y.F., P.H. and P.B. conceived the experiment. Y.F and J.K took data. Y.F, J.K. and L.M analyzed the results. Y.F. and Z.Z. constructed the analytical model and performed simulations. Y.F. and P.B. wrote the manuscript with input from all authors.

**Data availability:** The data supporting the findings of this study are available from the corresponding author upon request.

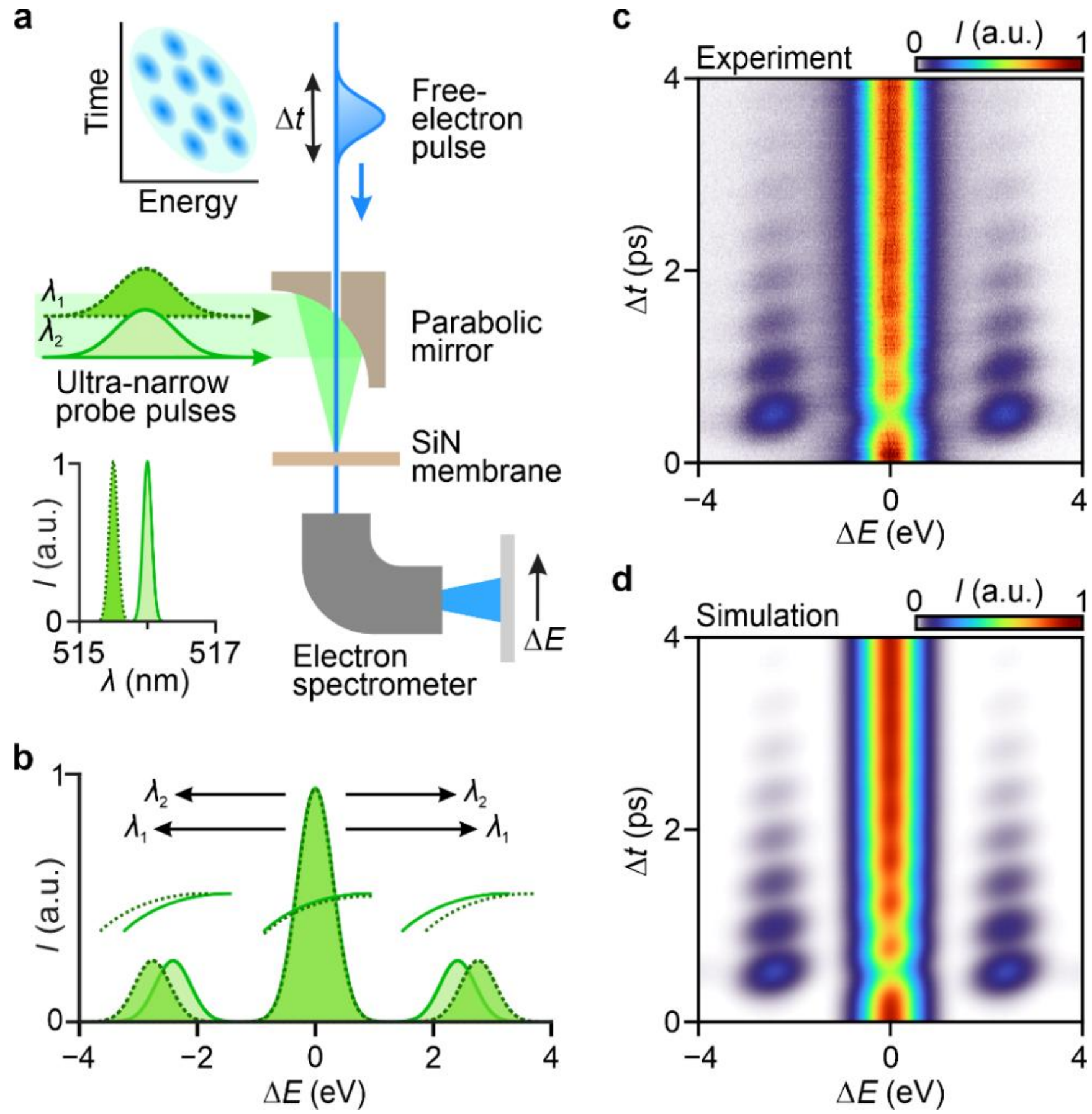

**Fig. 1. Concept of free-electron spectral shearing interferometry. a**, Experimental scenario. Quantum states of free-electron pulses are measured by interacting with a spectral shearing probe laser. The probe laser is composed of two collinear optical pulses with ultra-narrow bandwidths (< 0.2 nm) and slightly different central wavelengths (bottom left inset). It is tightly focused to interact with an unknown free electron ensemble (top left inset) in the vicinity of a silicon nitride membrane. The laser-modulated electron spectrum is recorded as a function of the time delay $\Delta t$ between the electron state and the laser wave to extract the free-electron density matrix. **b**, Principle of our spectral shearing interferometry. Two quasi-monochromatic laser waves induce two coherent quantum paths into positive and negative energy sidebands, letting the original electron wavefunction interfere with its duplicate at an energetic shift. Filled shape, spectrum; floating curves, spectral phases $\arg[\psi(\varepsilon)]$. Dotted green curves refer to the $\lambda_1$ quantum path and solid green curves denote the $\lambda_2$ quantum path. **c**, Measured free-electron interferogram as a function of $\Delta t$, acquired with probe lasers of $\lambda_1 = 514$ nm and $\lambda_2 = 515.9$ nm. **d**, Simulation result.

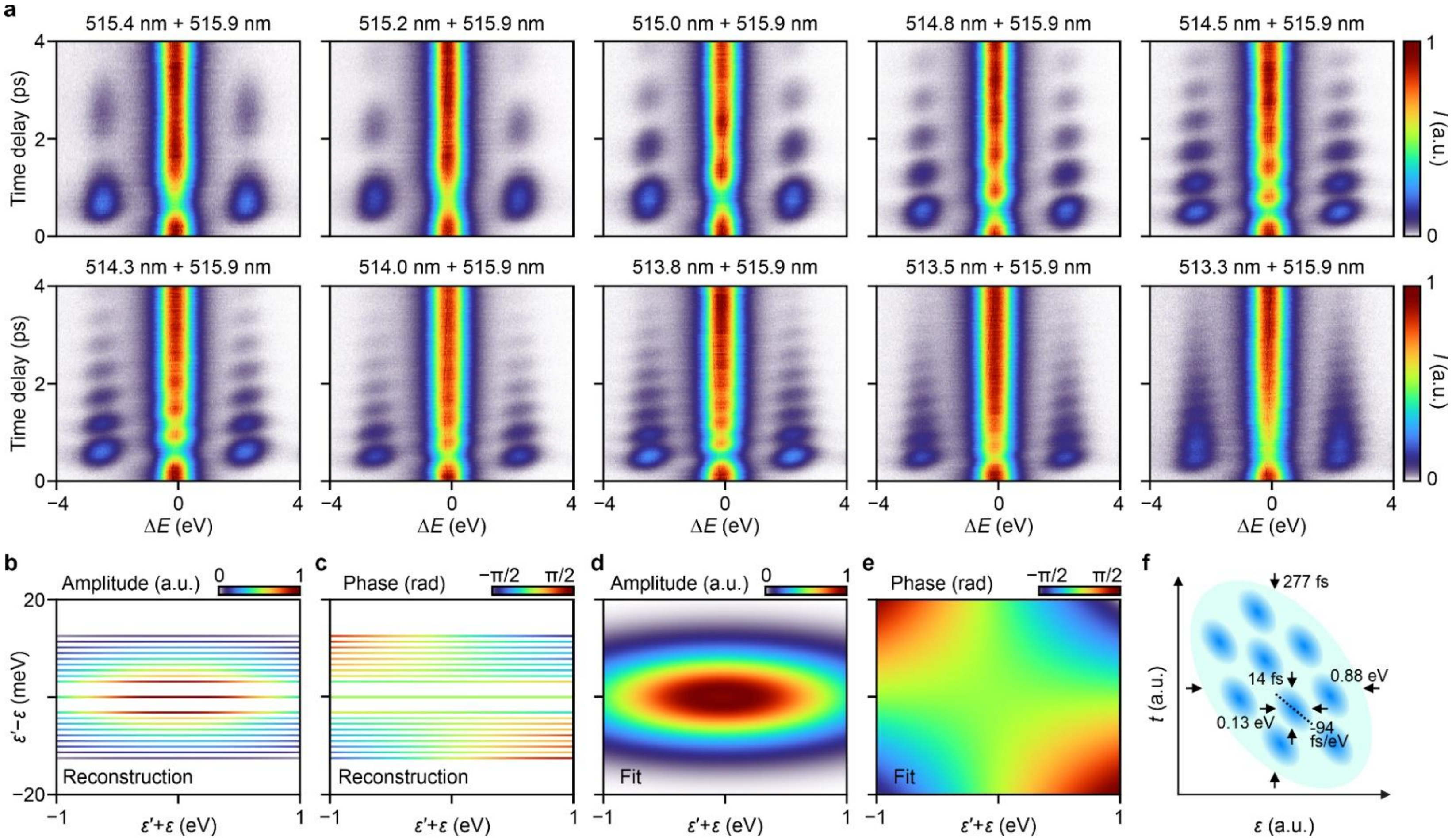

**Fig. 2. Free-electron spectral shearing interferometry results and the reconstructed density matrix. a**, Measured laser-modulated free-electron interferograms as a function of the electron-laser time delay (vertical axis) for various wavelengths of the synthesized laser wave. From left to right, upper to lower row, we have $\Delta E_{\mathrm{ph}}$ = 2.3 meV, 3.2 meV, 4.2 meV, 5.4 meV, 6.5 meV, 7.7 meV, 8.9 meV, 10.1 meV, 11.2 meV and 12.4 meV. **b**, Measured amplitude of the density matrix, obtained from the measured interferograms. White areas correspond to the regions that are not covered by experiment. **c**, Measured phase of the density matrix. **d**, Fitted amplitude of the density matrix. **e**, Fitted phase of the density matrix. **f**, Depiction of the measured mixed state in phase space. Light blue denotes the ensemble and dark blue denotes the individual pure states. The numbers shown are best fit results characterizing the individual wavepackets as well as the ensemble (see text).

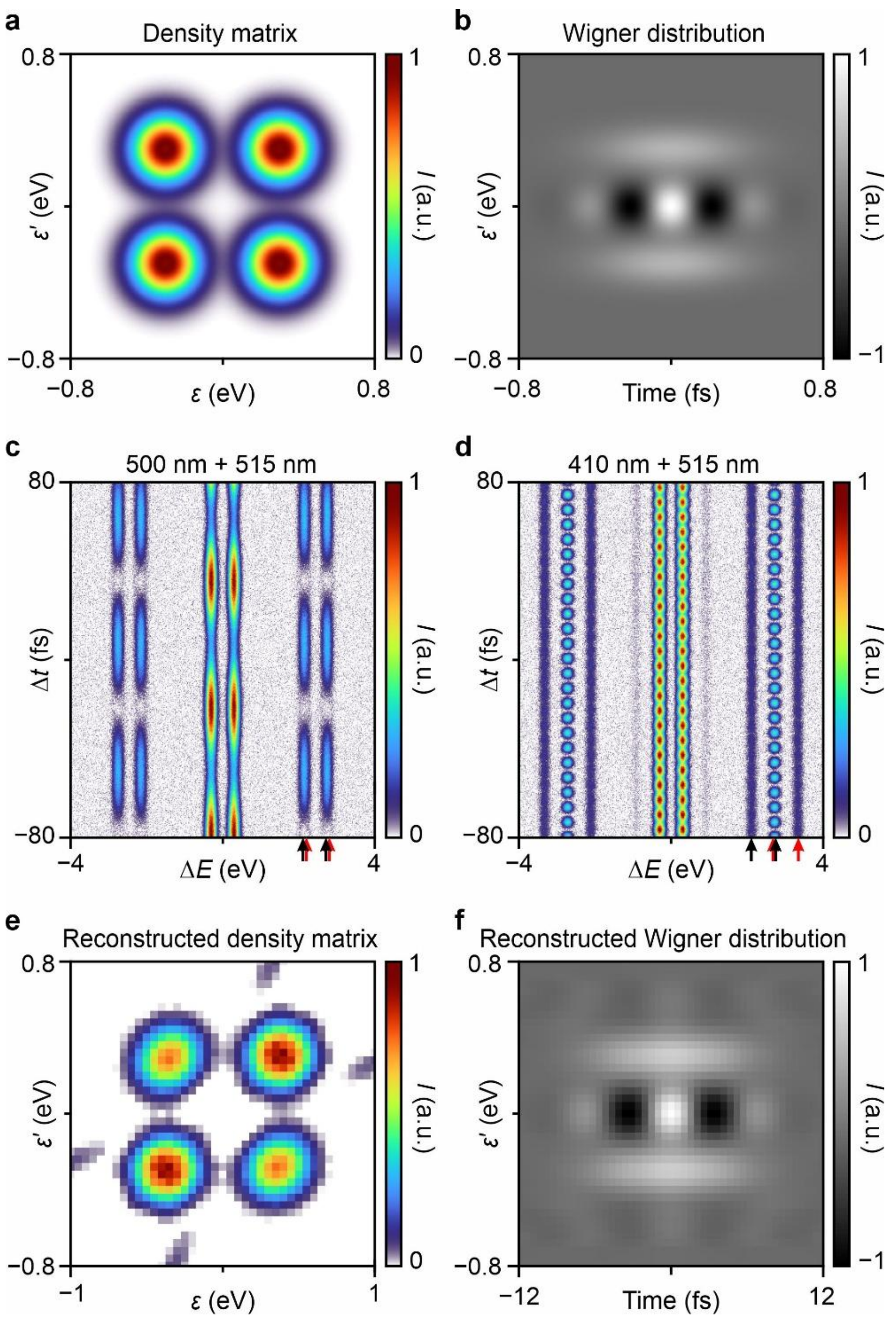

**Fig. 3. Simulated quantum state tomography of a free-electron cat state. a**, Calculated density matrix of a quantum state formed by superposition of two pure states with different central kinetic energies (180 keV ± 0.3 eV)**. b**, Corresponding Wigner distribution, showing characteristic negative interferences of a cat state. **c**, Simulated experimental spectral shearing data for a wavelength pair of 500 nm and 515 nm. **d**, Simulated experimental data for a wavelength pair of 410 nm and 515 nm. Black and red arrows denote the energetic positions of the first order sidebands from the two laser waves, respectively. See Extended Data Figure 4 for more details. **e**, Reconstructed density matrix. **f**, Reconstructed Wigner distribution. Although we apply simulated noise, we can very well reconstruct all relevant features of the original cat state.

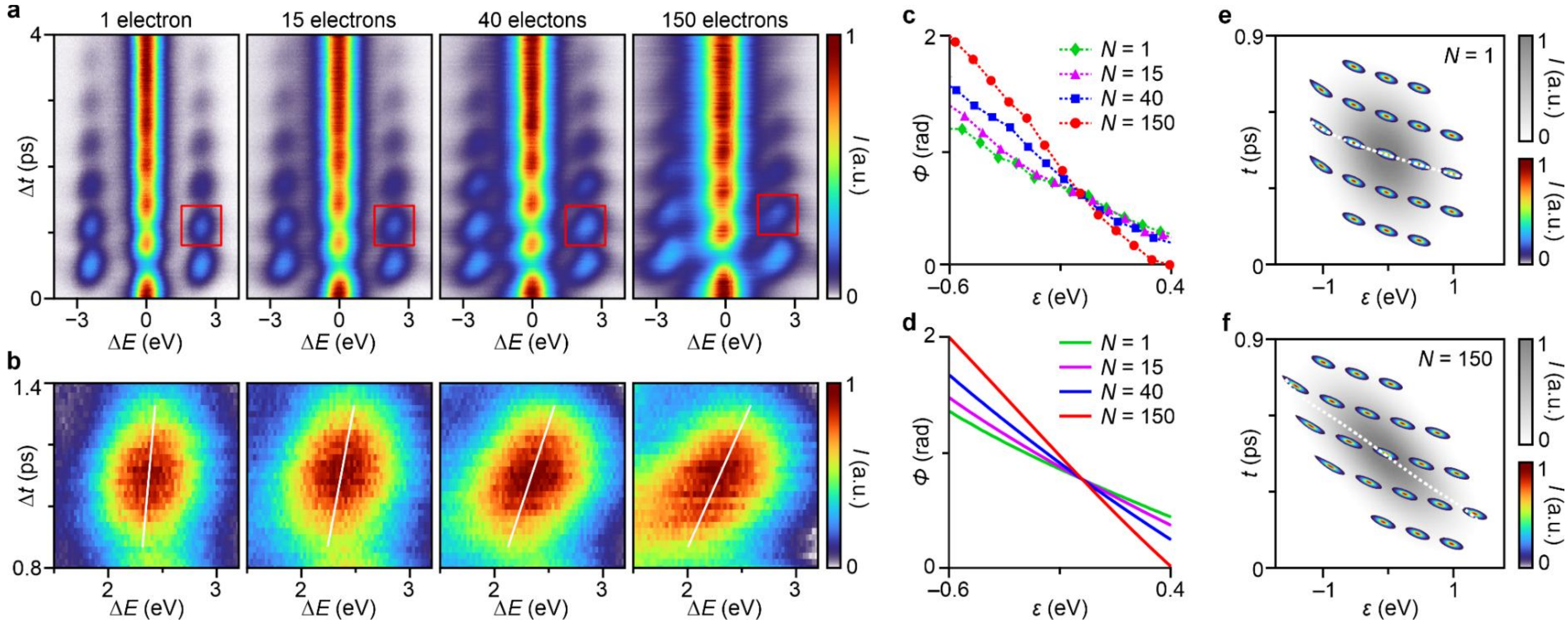

**Fig. 4. Multi-electron effects.** **a**, Measured free-electron interferograms for more and more electrons per pulse. **b**, Zoom into the tilted sideband features in these interferograms (red squares in **a**). White solid lines show the peak shift with respect to $\Delta t$. **c**, Measured density matrix phase $\Phi(\varepsilon)$ as a function of the number of electrons per pulse $N$. **d**, Simulated $\Phi(\varepsilon)$. **e**, Simulated phase-space distribution for the electron ensemble (grey) and selected pure wavepackets (color) in the absence of Coulomb interactions. **f**, Simulated phase-space distribution for 150 electrons in the cloud. The white dashed lines indicate the chirp of the electron ensemble. Individual electrons remain surprisingly coherent in energy and time.

## Methods

**Experiment:** We use an ultrafast transmission electron microscope with a W/ZrO Schottky field emitter (JEM F200, JEOL) at an electron energy of 180 keV and a femtosecond laser (Carbide, LightConversion) with 40-W output power, 1030-nm central wavelength, 250-fs pulse duration, and 2-MHz repetition rate. Extended Data Fig. 1 shows the experimental setup. We frequency-double the laser pulses using a β-barium borate (β-BBO) crystal and then split the beam into two paths. One path is used for photoemission of femtosecond electron pulses [31]. In this path, the laser power is set to be 100 μW for the results of Figs. 1 and 2, and 100 μW, 200 μW, 300 μW, 500 μW for Fig. 3. From the resulting 1-150 electrons per pulse, only one of them is transmitted through our microscope's apertures to the diagnostics part. The other laser path is used to generate the spectral shearing light fields in an optical interferometer. In each of the arms, we place an ultra-narrow bandwidth optical filter (Andover), whose nominal bandwidth is 0.15 nm (full width at half maximum, FWHM), to select specific wavelengths for the interaction. The transmission wavelength of one filter is fixed at 515.9 nm and the other one is rotated to finely adjust the wavelength detuning. Rigid and compact construction provides a passive phase stability. Resulting spectra from the interferometer are recorded by an optical spectrum analyzer (Yokogawa); see Extended Data Fig. 2. Using a half-wave plate before the interferometer, the output power of each arm is set to 200 μW. The combined light wave is then overlapped with the electron beam using a parabolic mirror, producing a spot with a radius of ~4 μm on a 50-nm-thick SiN membrane (Norcada). At the laser focus, the optical wave has side lobes with an electric field with z-component [48] that provide a longitudinal momentum modulation for the electrons, that is, energy gain and loss. The electron beam diameter is <50 nm (FWHM) and thus much smaller than the optical spot on the membrane, providing a spatially homogeneous modulation over the whole electron beam. Electron spectra are obtained with a magnetic post-column electron energy analyzer (CEFID, CEOS) and an event-based direct electron detector (Timepix3, Amsterdam Scientific Instruments), which can measure every single electron [49]. Only single-electron events are recorded. The integration time is 10 s for each electron spectrum, resulting in ~20 minutes for each delay-scanned interferogram.

**Temporal response of ultra-narrow optical filters**: In our Fabry-Perot filters, the incoming femtosecond laser pulses undergo multiple reflections while releasing a small portion of laser wave each time. The temporal envelope of the output laser wave thus becomes an exponentially modified Gaussian function:

$$f(t)=\sum_{n=0}^{\infty} r^n \exp\left[-\frac{(t-n\delta t)^2}{2\sigma_{\text{en}}^2}\right]$$
$$\approx (\text{Gaussian}) * (\text{Exponential}) = \exp\left(\frac{\sigma_{\text{en}}^2}{2\tau^2}-\frac{t}{\tau}\right)\text{erfc}\left(\frac{\sigma_{\text{en}}^2/\tau - t}{\sqrt{2}\sigma_{\text{en}}}\right) \quad (2)$$

where $r$ is the field-amplitude survival factor after each round trip of the Fabry–Perot cavity, $\delta t$ determines the center of each Gaussian envelope, $\sigma_{\text{en}}$ is the standard deviation of each Gaussian envelope and $\tau = -\delta t/\ln r$ is a decay time of the exponential term. In our case, we have $\sigma_{\text{en}} = 100$ fs and $\tau = 1.5$ ps. This physics explains the general shapes of the results in Figs. 1-3 on picosecond time scales.

**Simulations of interferograms**: To simulate the data of Fig. 1d, we neglect realistic acceleration processes and multi-electron effects in the electron microscope. We assume a partially incoherent state in time and energy. We first solve the relativistically modified Schrödinger equation for each pure state, $i\hbar\partial_t\psi(z', t) = \hat{H}\psi(z', t)$ [50], where $\psi(z', t)$ is a pure-state electron wavefunction in the local coordinate around the center of electron wavepacket. $E_{\text{el}} = 180$ keV is the central energy and $\hat{H}$ is the Hamiltonian. The wavefunction after laser-modulation is given by $\psi(z') = \psi_0(1+i\xi L_{\text{D}})^{-1/2}\exp\{-z'^2/[4v_{\text{el}}^2\sigma_t^2(1+i\xi L_{\text{D}})]\}\exp\{if(t)|g_1|\sin[\omega_1/v_{\text{el}}z'+\arg(g_1)]\}\exp\{if(t)|g_2|\sin[\omega_2/v_{\text{el}}z'+\arg(g_2)]\}$, where $f(t)$ is given by Eq. (2), $\psi_0$ is a normalization coefficient, $\sigma_t$ is the temporal coherence length, $L_{\text{D}}$ is the effective dispersive propagation in free space before interaction, $\xi$ is given by $\xi = \hbar/(2m_{\text{el}}\gamma^3\sigma_z^2 v_{\text{el}})$, $m_{\text{el}}$ is the electron mass, $\gamma$ is the Lorentz factor, $v_{\text{el}}$ is the electron velocity, $\omega_1$ and $\omega_2$ are the angular frequency for $\lambda_1$ and $\lambda_2$, respectively, and $g_1$ and $g_2$ is the electron-laser coupling strength for $\lambda_1$ and $\lambda_2$, respectively. The spectrum of the electron as a function of $t$ is obtained by performing a Fourier transformation on $\psi(z')$. We use the measured $\sigma_t = 6$ fs (Fig. 2f) and assume $L_{\text{D}} = 10$ m to match the measured electron chirp of -94 fs/eV (Fig. 2f). Then, we consider a Gaussian energy jittering and temporal jittering of the pure state, using the measured values of $\sigma_{\varepsilon 0} = 0.37$ eV and $\sigma_{t0} = 118$ fs, respectively. The peak field strengths of the two laser waves are 0.05 V/nm. We note that $L_{\text{D}} = 10$ m, which is much larger than the realistic length scale of our microscope, reflects that most of dispersion of our electrons is accumulated during the acceleration process in the microscope.

**Analytical density matrix of femtosecond electrons**: We derive an analytical expression for the density matrix that can be expected for electrons that are photo-emitted by femtosecond laser pulses from a tungsten needle tip. Temporal jitter comes from random emission in the laser pulse duration.

Energy jitter is caused by surface effects, a spatially inhomogeneous Schottky effect, and/or fluctuations of the acceleration voltage. For a coherent free-electron wavepacket with chirp, its wavefunction ansatz at time $t = 0$ is

$$\psi(z) = \frac{1}{[2\pi(1+C^2)]^{1/4}} \exp\left[ -\frac{(z-z_0)^2}{4(1+iC)\sigma_z^2} + iq_0 z + i\varphi_0 \right], \tag{3}$$

where $C$ is a linear chirp parameter, $q_0$ is the central momentum, $z_0$ is the coordinate of pulse center, $\sigma_z$ is the standard deviation of electron pulse length, and $\varphi_0$ is a global phase. Its wavefunction in momentum space is

$$\tilde{\psi}(q) = \left[ \frac{1}{2\pi\sigma_q^2} \frac{(1+iC)^2}{1+C^2} \right]^{1/4} \exp\left[ -\frac{(q-q_0)^2}{4\sigma_q^2}(1+iC) - i(q-q_0)z_0 + i\varphi_0 \right], \tag{4}$$

where $\sigma_q = 1/(2\sigma_z)$ is the standard deviation of electron momentum. The density matrix for a coherent wavepacket is expressed as $\rho = |\tilde{\psi}\rangle\langle\tilde{\psi}|$. Each density matrix element in momentum space is given by

$$\begin{aligned} \rho(q_1,q_2) &= \tilde{\psi}(q_1)\tilde{\psi}^*(q_2) \\ &= \frac{1}{\sqrt{2\pi\sigma_q^2}} \exp\left[ -\frac{[q_0-(q_1+q_2)/2]^2}{2\sigma_q^2} - \frac{(q_1-q_2)^2}{8\sigma_q^2} + iC\frac{[q_0-(q_1+q_2)/2](q_1-q_2)}{2\sigma_q^2} - i(q_1-q_2)z_0 \right]. \end{aligned} \tag{5}$$

We take the temporal and energy jittering into consideration by assuming that the statistical distribution for $(q_0, z_0)$ is a chirped Gaussian. The covariance matrix takes the following form:

$$\Sigma = \begin{bmatrix} \sigma_{q_0}^2 & \sigma_{qz} \\ \sigma_{qz} & \sigma_{z_0}^2 \end{bmatrix}. \tag{6}$$

The off-diagonal term $\sigma_{qz}$ accounts for a linear chirp of the incoherent distribution, and the determinant of $\Sigma$ is $\det(\Sigma) = \sigma_{q0}^2\sigma_{z0}^2 - \sigma_{qz}^2$. Thus, the statistical distribution for $(q_0, z_0)$ is

$$f_s(q_0,z_0) = \frac{1}{2\pi\sqrt{\sigma_{q_0}^2\sigma_{z_0}^2 - \sigma_{qz}^2}} \exp\left[ -\frac{\sigma_{z_0}^2(q_0-q_c)^2 - 2\sigma_{qz}(q_0-q_c)(z_0-z_c) + \sigma_{q_0}^2(z_0-z_c)^2}{2(\sigma_{q_0}^2\sigma_{z_0}^2 - \sigma_{qz}^2)} \right], \tag{7}$$

where $q_c$ and $z_c$ are the mean values for $q_0$ and $z_0$. The ensemble average of the elements of the density matrix is

$$\begin{aligned} \langle \rho(q_1,q_2) \rangle &= \int dq_0 \int dz_0 f_s(q_0,z_0)\rho(q_1,q_2) \\ &= \frac{1}{2\pi\sqrt{\sigma_q^2+\sigma_{q_0}^2}} \exp\left\{ -\frac{(q_1-q_2)^2}{8\sigma_q^2} - \frac{[(q_1+q_2)/2-q_c]^2}{2(\sigma_q^2+\sigma_{q_0}^2)} - \frac{[\sigma_{z_0}^2\sigma_q^2 + \det(\Sigma) - \sigma_{qz}C](q_1-q_2)^2}{2(\sigma_q^2+\sigma_{q_0}^2)} \right. \\ &\quad \left. -i(q_1-q_2)\left[ z_c + \frac{(2\sigma_{qz}+C)[(q_1+q_2)/2-q_c]}{2(\sigma_q^2+\sigma_{q_0}^2)} \right] \right\} \end{aligned} \tag{8}$$

Converting equation (8) into energy-time space ($\varepsilon$, $t$) by following $q = \varepsilon/(\hbar v_{\mathrm{el}})$ and $z = -v_{\mathrm{el}}t$ one obtains the final expression for the density matrix of free electrons,

$$\begin{aligned}\langle\rho(\varepsilon_1,\varepsilon_2)\rangle = \frac{1}{2\pi\sqrt{\sigma_\varepsilon^2+\sigma_{\varepsilon_0}^2}}\exp\Bigg\{&-\frac{(\varepsilon_1-\varepsilon_2)^2}{8\sigma_\varepsilon^2}-\frac{[(\varepsilon_1+\varepsilon_2)/2-\varepsilon_c]^2}{2(\sigma_\varepsilon^2+\sigma_{\varepsilon_0}^2)}-\frac{[\sigma_{t_0}^2\sigma_\varepsilon^2+\det(\Sigma)-\sigma_{\varepsilon t}C](\varepsilon_1-\varepsilon_2)^2}{2\hbar^2(\sigma_\varepsilon^2+\sigma_{\varepsilon_0}^2)}\\ &-i\frac{\varepsilon_1-\varepsilon_2}{\hbar}\left[t_c+\frac{(2\sigma_{\varepsilon t}+C)[(\varepsilon_1+\varepsilon_2)/2-\varepsilon_c]}{2(\sigma_\varepsilon^2+\sigma_{\varepsilon_0}^2)}\right]\Bigg\}\end{aligned} \tag{9}$$

Equation (9) shows that the parameters for the coherent electron wavepacket ($\sigma_\varepsilon$ and $C$) and the parameters for the incoherent distribution ($\sigma_{\varepsilon 0}$, $\sigma_{t0}$ and $\sigma_{\varepsilon t}$) can be extracted separately from the measured density matrix. The aspect ratio $\eta$ of the amplitude in the coordinate ($\varepsilon' - \varepsilon$, $\varepsilon' + \varepsilon$) scales the degree of decoherence of the electron ensemble and it is given by

$$\eta = \left[\frac{\sigma_\varepsilon^2+\sigma_{\varepsilon 0}^2}{\sigma_\varepsilon^2}+\frac{4(\sigma_{t0}^2\sigma_\varepsilon^2+\det(\Sigma)-\sigma_{\varepsilon t}C)}{\hbar^2}\right]^{-1/2}. \tag{10}$$

**Theory of free-electron quantum state tomography**: We consider a density matrix $\rho$ of free-electron pulses. The time-dependent Hamiltonian corresponding to the electron-photon interaction includes an unperturbed part $H_0$ and an interaction part $H_{\mathrm{I}}$. After interaction, the probability to find an electron in a final state with energy $\varepsilon_{\mathrm{f}}$ is given by $S(\varepsilon_{\mathrm{f}}, \Delta t, \Delta E_{\mathrm{ph}}) = \langle\varepsilon_{\mathrm{f}}|U_{\mathrm{I}}(t, \Delta t, \Delta E_{\mathrm{ph}})\rho U_{\mathrm{I}}^\dagger(t, \Delta t, \Delta E_{\mathrm{ph}})|\varepsilon_{\mathrm{f}}\rangle$, where $U_{\mathrm{I}}$ is a time-ordered unitary operator in the interaction picture,

$$U_I(t,\Delta t,\Delta E_{ph}) = T\left[\exp\left(-i/\hbar\int_{-\infty}^{t}dt' H_{\mathrm{I}}'(t',\Delta t,\Delta E_{ph})\right)\right]. \tag{11}$$

Here, $T$ is the time-ordering operator, and the interaction Hamiltonian $H_{\mathrm{I}}'$ is derived by analytically solving time-dependent relativistically modified Schrödinger equation, giving rise to an expression in the energy basis

$$H_{\mathrm{I}}' = e^{iH_0t/\hbar}H_{\mathrm{I}}e^{-iH_0t/\hbar} \propto e^{iH_0t/\hbar}\left[\int dE\int d\varepsilon\cdot g\sum_{m=1}^{2}f_m(t)\left(e^{i\omega_m(t-\Delta t)}+e^{-i\omega_m(t-\Delta t)}\right)|E\rangle\langle\varepsilon|+\mathrm{H.c.}\right]e^{-iH_0t/\hbar} \tag{12}$$

where $g$ is the coupling strength between electrons and photons, $m$ labels the two laser wavelengths, $\omega_m$ is the angular laser frequency, $f$ is the exponentially modified Gaussian envelope of laser pulses according to Eq. (2). In the derivation of Eq. (12), we have assumed that the coupling strength is identical for the two wavelength components. Using Eq. (12) and the rotating wave approximation, we obtain an explicit perturbative expansion for $U_{\mathrm{I}}$,

$$U(t,\Delta t,\Delta E_{ph}) \approx 1-\frac{i}{\hbar}\sum_{m=1}^{2}\int\int dEd\varepsilon\cdot\left[\frac{g}{1-re^{i(\omega_{E\varepsilon}-\omega_m)\delta t}}e^{-\sigma_{\mathrm{en}}^2(\omega_{E\varepsilon}-\omega_m)^2/2}e^{i\omega_{E\varepsilon}\Delta t}|E\rangle\langle\varepsilon|+H.c.\right], \tag{13}$$

where $\omega_{E\varepsilon}\hbar$ is the energy difference between the final state of electrons and the initial state. If the final state of electrons does not overlap with its initial state, one will get

$$S(\varepsilon_{\mathrm{f}},\Delta t,\Delta E_{\mathrm{ph}}) = -\left\langle \varepsilon_{\mathrm{f}} \right| \frac{i}{\hbar}\sum_{m=1}^{2}\mathrm{X}\cdot\rho\cdot\frac{i}{\hbar}\sum_{n=1}^{2}\mathrm{X}\ \left|\varepsilon_{\mathrm{f}}\right\rangle. \tag{14}$$

Here, X refers to the terms inside the summation of $m$ in Eq. (13). If the values of the density matrix elements are almost homogeneous within the bandwidth of the probe laser pulses, the envelope of the laser wave inside Eq. (14) can be simplified to

$$S(\varepsilon_{f},\Delta t,\Delta E_{\mathrm{ph}}) = f^{2}(\Delta t)\,|\,g\,|^{2}\left\{\rho(\varepsilon,\varepsilon)+\rho(\varepsilon',\varepsilon')+2\,\mathrm{Re}\left[e^{i\Delta E_{\mathrm{ph}}\Delta t/\hbar}\rho(\varepsilon,\varepsilon')\right]\right\}. \tag{15}$$

Here, the accessible final state and the initial state are determined by $\varepsilon = \varepsilon_f - \omega_1\hbar$, $\varepsilon' = \varepsilon_f - \omega_2\hbar$ and $\varepsilon' - \varepsilon = \Delta E_{\mathrm{ph}}$. An application of Eq. (15) for atomic cases can be found in Ref. [51].

Equation (15) is obtained by making three main assumptions: (1) The two wavelength components induce the same coupling strength, (2) the initial state of the electrons is energetically well separated from the ±1 sidebands in the spectrum, and (3) the bandwidth of all laser waves is small compared to the energy scale over which the electron density matrix varies. In our experiment, we ensured these conditions by (1) balancing the measured electron sideband intensities generated by each individual laser wave, (2) using the second harmonic of our ytterbium laser to produce high-energy sidebands, and (3) applying an optical filter with a bandwidth below 0.2 nm.

**Reconstruction of the density matrix**: The diagonal elements are directly read from the measured electron spectrum without any laser modulation. The off-diagonal elements are extracted by performing a two-dimensional fit to the experimental data (Extended Data Fig. 3). We use a summation of three Voigt profiles to fit the -1st sideband, zero-loss peak and +1st sideband along the energy axis (Extended Data Fig. 3b) for each time delay (Extended Data Fig. 3c). Along the time-delay axis, we use a function in the form of

$$F(t) = f^{2}(t)[P_1 + P_2\cos(P_3 t + P_4)]. \tag{16}$$

to fit the data at each $\Delta E$ (Extended Data Fig. 3d). Comparing with Eq. (15), the fitting parameter $P_2$ describes the amplitude $|\rho(\varepsilon,\varepsilon')|$ and the parameter $P_4$ describes the phase $\arg[\rho(\varepsilon,\varepsilon')]$. The coordinate of the extracted value in the density matrix is given by $(\varepsilon, \varepsilon')$ (dark blue square in Extended Data Fig. 3e). For the ±1th sideband, $\varepsilon$ and $\varepsilon'$ are determined by $\Delta\varepsilon = \Delta E \mp E_{\mathrm{ph}}$ and $\varepsilon' = \varepsilon + \Delta E_{\mathrm{ph}}$. Repeating the fitting for each $\Delta E$ yields the density matrix elements along the light blue squares in Extended Data Fig. 3e. Then, tuning the wavelength difference $\Delta E_{\mathrm{ph}}$ of the probe laser, we reconstruct the discrete density matrix of Fig. 2b-c; see also Extended Data Fig. 3f.

For closing the gaps (Fig. 2d-e), we use a maximum-likelihood estimation to reconstruct the density matrix in a continuous basis. We apply a fixed time jittering of 277 fs in Eq. (9) and set a prior value of $8\times10^{-3}$ with a tolerance range 20% for $\eta$ to avoid unreasonable local minima. We also assume that the incoherent chirp is equal to the coherent chirp, that is, $C/2\sigma_\varepsilon^2 = \sigma_{\varepsilon t}/\sigma_{\varepsilon 0}^2$. Under these assumptions, we fit three central parameters ($\sigma_\varepsilon$, $\sigma_{\varepsilon 0}$, and $C$) and two auxiliary parameters ($t_c$ and $\varepsilon_c$) in Eq. (9) to the measured discrete density matrix of Fig. 2b-c. In this way, we obtain the continuous density matrix of Fig. 2d-e as well as the fitting parameters shown in Fig. 2f.

**Simulation of cat-state tomography**: The wavefunction of the assumed free-electron cat state is given by

$$|\psi(t)\rangle = \left(2\pi\sigma_t^2\right)^{-1/4} e^{-t^2/4\sigma_t^2} e^{-iE_{el}t/\hbar} \left(e^{i\Delta Et/\hbar} + e^{-i\Delta Et/\hbar}\right) \tag{17}$$

where $\sigma_t = 6$ fs, $E_0 = 180$ keV and $\Delta E = 0.3$ eV. We calculate the expected interferometry (Fig. 3 and Extended Data Fig. 4) by solving the relativistically modified Schrödinger equation as described above for a pure-state case. The wavelength of one of the probe lasers is scanned from 350 nm to 515 nm, while the other one is kept at 515 nm. The intensity of each probe laser wave is controlled to provide a coupling strength $g \approx 0.7$. We then add Gaussian readout noise with a standard deviation of 20 counts to each pixel in the simulated interferogram and mimic the finite experimental energy resolution by convolution with a Gaussian point-spread function with a standard deviation of two pixels. Due to this realistic noise, the peak amplitude of the reconstructed off-diagonal elements (Fig. 3e) is slightly reduced as compared to the original cat state (Fig. 3a). Nevertheless, the general Wigner distribution and negative values are clearly reproduced (Fig. 3f).

**Density matrix and phase space distributions with space charge**: To simulate how the density matrix changes in the presence of Coulomb effects, we start with the Schrödinger equation of the electron under investigation,

$$i\hbar \frac{\partial \psi(z,t)}{\partial t} = \left[ -\frac{\hbar^2}{2m_e} \nabla^2 + U_0(z,t) + U_c(z,t) \right] \psi(z,t). \tag{18}$$

Here, $U_0$ is the potential of the electrostatic field in the electron microscope and $U_c$ is the electrodynamic potential from other surrounding electrons. To derive a tractable approximation for the evolution of this Schrödinger equation with time, we write the wavefunction of the electron in a one-dimensional eikonal form, $\psi(z, t) = A(z, t)\exp[iS_{sc}(z, t)/\hbar]$, where $A$ is the slowly varying amplitude and $S_{sc}$ is the rapidly varying action. Substituting this ansatz into Eq. (18) and expanding in powers of $\hbar$ yields the Hamilton-Jacobi equation [52]:

$$\partial_t S_{sc} + \frac{(\partial_z S_{sc})^2}{2m_e} + U_0 + U_c = 0. \tag{19}$$

Equation (19) describes semi-classical properties of the Schrödinger equation: One can identify $\partial_z S_{sc} = p$ with $p$ being the classical momentum of the electron and $H(z, p, t) = p^2/2m_e + U_0 + U_c$ with $H$ being the classical Hamiltonian. The solution of Hamilton-Jacobi equation $z(t)$ satisfies the classical equations $dz/dt = p/m_e$ and $dp/dt = -\partial(U_0 + U_c)/\partial z$, which defines a semiclassical phase-space $(t, \varepsilon)$ mapping for the electron from an initial distribution at the tip to a final distribution in the sample plane [52]. Therefore, the final phase-space distribution of our investigated single-electron can be obtained by solving the hypothetical motions of a set of point-like parts of the electron, sampled from the initial phase-space distribution (Extended Data Fig. 5), similar to related approaches in attosecond physics [53,54]. This approximation is applicable to our experiment because the electron wavelength is small compared to the characteristic length scales of the electrodynamic field and the space-charge potential in the electron microscope.

In our electron microscope, the motion of the point-like electrons and parts of their wave function can be divided into four segments (Extended Data Fig. 5): (i) acceleration from the tip to the extractor, (ii) acceleration from the extractor to the first anode, (iii) acceleration from the first anode to the second anode and (iv) free propagation to the sample position. Between all stages, apertures block part of the electron beam, and interaction with the electron cloud is only considered in the region close to the emitter tip. We consider a three-dimensional electron cloud but simulate quantum effects only on the optical axis, the paths later selected for measurement. We solve the one-dimensional motion of the point-like parts of each electron in segment (i) with a fifth-order Runge-Kutta method. Motions in segments (ii), (iii) and (iv) are solved analytically. In all simulations, we consider relativistic effects but neglect the spin of the electrons.

To simulate the effects of the surrounding electron cloud to our single electrons, we apply a mean-field theory [39] and approximate the space charge of our 15-150 electrons as a continuous, three-dimensional, isotropic Gaussian charge density whose spatial spread $\sigma(t)$ in each dimension is described by a diffusion function $\sigma^2(t) = \sigma^2(0) + 2Dt$, using an initial spread of $\sigma(0) = 2$ μm, a value comparable to the diameter of our tip apex, and a diffusion rate $D = 2$ m/s [28]. The spatial integral of the charge density equals the total number of single electrons $N$. The time-dependent motion of the center of this charge density is determined by averaging two point-like trajectories that are not affected by Coulomb interactions. The emission times and initial energies of these two point-like paths are −350 fs, 1.2 eV and 350 fs, 1.2 eV, respectively. The electric field of the moving and spreading charge density is then calculated by Gauss's law, neglecting magnetic fields.

For the simulation of one quantum-mechanical electron inside of this moving cloud, on the optical axis, we sample its wave function and semi-classically propagate each part as described above [53,54]. We assume coherent states with standard deviations $\sigma_\varepsilon = 0.06$ eV and $\sigma_t = 6$ fs (similar to the measured pure width of 0.13 eV FWHM and coherence length of 14 fs FWHM, respectively). We solve the point-like motions of these parts of the electron wave function as introduced above [53,54]. The weight of each path is given by the product of these two Gaussian distributions. Then, we collect the arrival times and energies of all point-like paths at the location of measurement, ~65 cm after the emitter tip (Extended Data Fig. 5). In this way, we obtain the phase-space distribution of one electron wavepacket.

Next, we consider jitter of the pure electron wavepackets in time and energy. We repeat the above single-electron simulations by injecting a regular array for the initial energy $\varepsilon_0$ and emission time $t_0$. We use for $\varepsilon_0$ a Gaussian distribution with a standard deviation $\sigma_{\varepsilon 0} = 0.37$ eV at a mean energy of 1.2 eV [31], and use for $t_0$ a Gaussian distribution with a standard deviation $\sigma_{t0} = 118$ fs (or 277 fs FWHM) at a mean time of 0 fs. The weight of each electron wavepacket is determined by the product of these two Gaussian distributions. The summation of all such electron wavepackets in phase space produces the incoherent ensemble of the electron wavepackets as shown in the grey intensity map of Fig. 4e and 4f; see also Extended Data Fig. 6. The colourful data in Fig. 4e and 4f and Extended Data Fig. 6 shows the phase space of each simulated pure state after propagation in the space charge cloud. The final density matrix can be approximated by Fourier transforming the phase-space distribution along the time axis, or it can be calculated rigorously from its definition by explicitly using the simulated electron amplitudes and phases.

Here we calculate the phase by solving Feynman path integral. Specifically, the phase $\phi = S_{sc}/\hbar + Et_{sam}/\hbar$ is given in terms of the classical action $S_{sc}$ which is calculated by integrating the Lagrangian of the point-like path between the tip and the sample plane [40,41], as described above. The term $Et_{sam}/\hbar$ accounts for the Legendre transform, where $E = \gamma_{sam} m_e c^2$ is the energy of the point-like part arriving at the sample plane, $t_{sam}$ is the arrival time at the sample plane, and $\gamma_{sam}$ is the Lorenz factor. For the simulation, we sample $\varepsilon$ for the initial energies of point-like paths in each electron wavepacket in a same way as described above, while fixing the initial time at $t = t_0$. This approximation avoids the substantial computational overhead arising from the rapidly oscillating free-electron phase. In this way, we obtain the electron wavefunction $\psi(\varepsilon)$ including a phase and amplitude, which directly provides the density matrix for a pure state of electron by $\rho_0(\varepsilon, \varepsilon') = \psi(\varepsilon)\psi^*(\varepsilon')$. Then we also sample $\varepsilon_0$ and $t_0$ for the pure state and we obtain the realistic density matrix for the mixed state of the electrons, $\rho(\varepsilon, \varepsilon') = \langle \rho_0(\varepsilon, \varepsilon') \rangle$. By performing a line cut of the phase of the full density matrix at $\varepsilon' - \varepsilon = 6.5$ meV, we obtain the one-dimensional results shown in Fig. 4d. Comparing the simulation results with the measured data (Fig. 4c), we identify the effective electron numbers in the electron cloud as $N = 1$, 15, 40 and 150 (Fig. 4d).

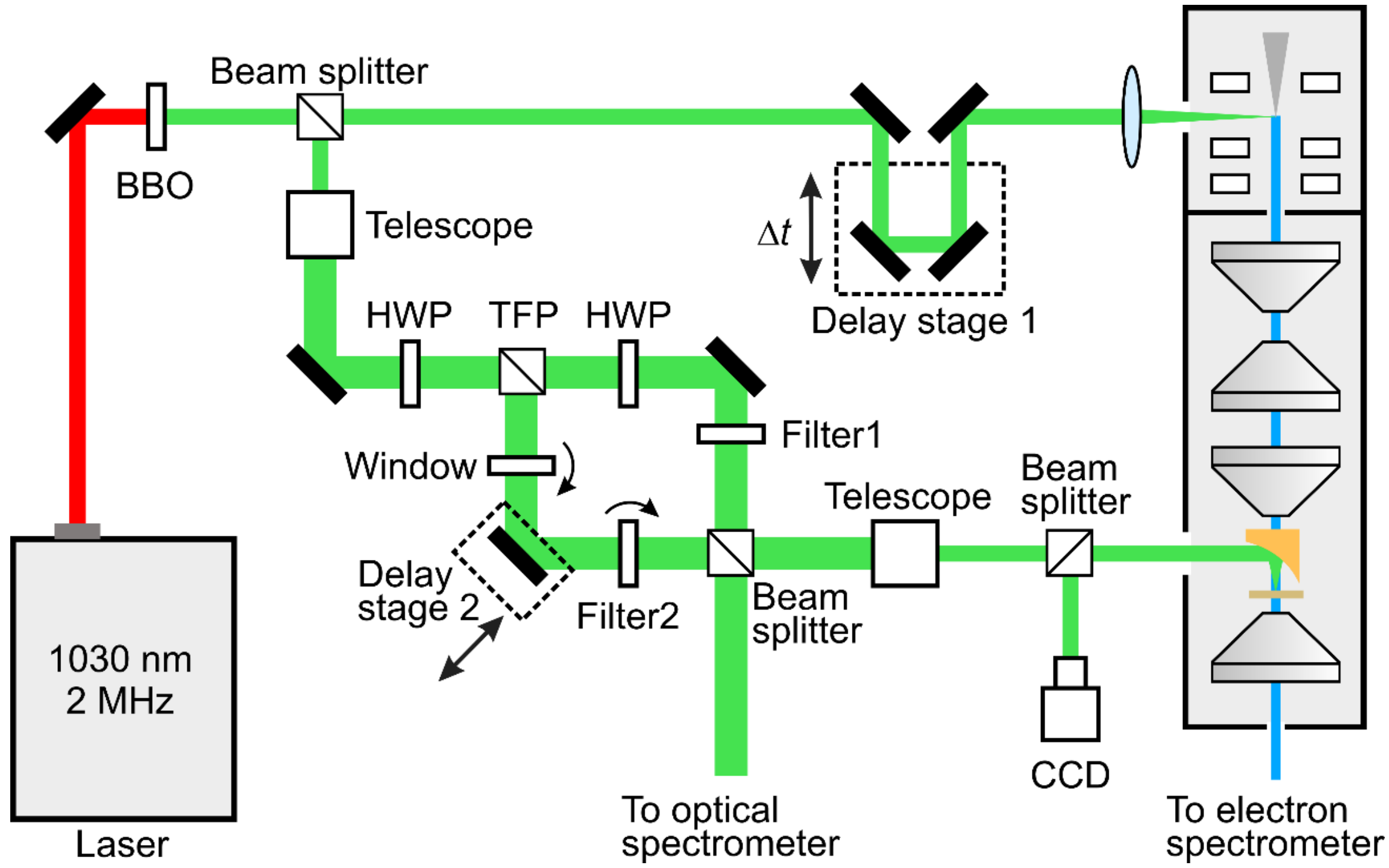

**Extended Data Fig. 1 | Experimental setup.** The 1030-nm fundamental laser beam (red) is frequency-doubled to 515 nm (green) in a β-barium borate crystal (BBO)and split into two paths. One beam triggers photoelectrons with a time delay $\Delta t$ (delay stage 1) and produces a pulsed electron beam (blue). The other beam passes a telescope and enters a compact Mach-Zehnder interferometer, where the beam is split by a thin-film polarizer (TFP) and recombined by a non-polarizing beam splitter. A half-wave plate (HWP) adjusts the relative beam powers. In each arm, we place a hyperfine narrowband filter. In the arm of Filter 1, a HWP rotates the polarization to match that of the other arm. A glass block (window) and delay (Delay stage 2) are included to provide equal propagation times in both arms. Filter 1 selects a fixed wavelength of 515.9 nm and Filter 2 can be tuned by rotation. At the output of the interferometer, one beam is sent to an optical spectrometer for diagnostics, while the other one is focused with an off-axis parabolic mirror (yellow) to interact with the electron pulses at a silicon nitride membrane (dark yellow) inside the electron microscope. A charge-coupled device (CCD) camera monitors the back-reflected beam from the sample. The modulated electron spectrum is recorded by an electron spectrometer. Black rectangles, optical mirrors; light blue ellipse, optical lens; grey small rectangles, electrostatic acceleration elements; grey triangle, electron emission tip; grey trapezoids, magnetic lenses.

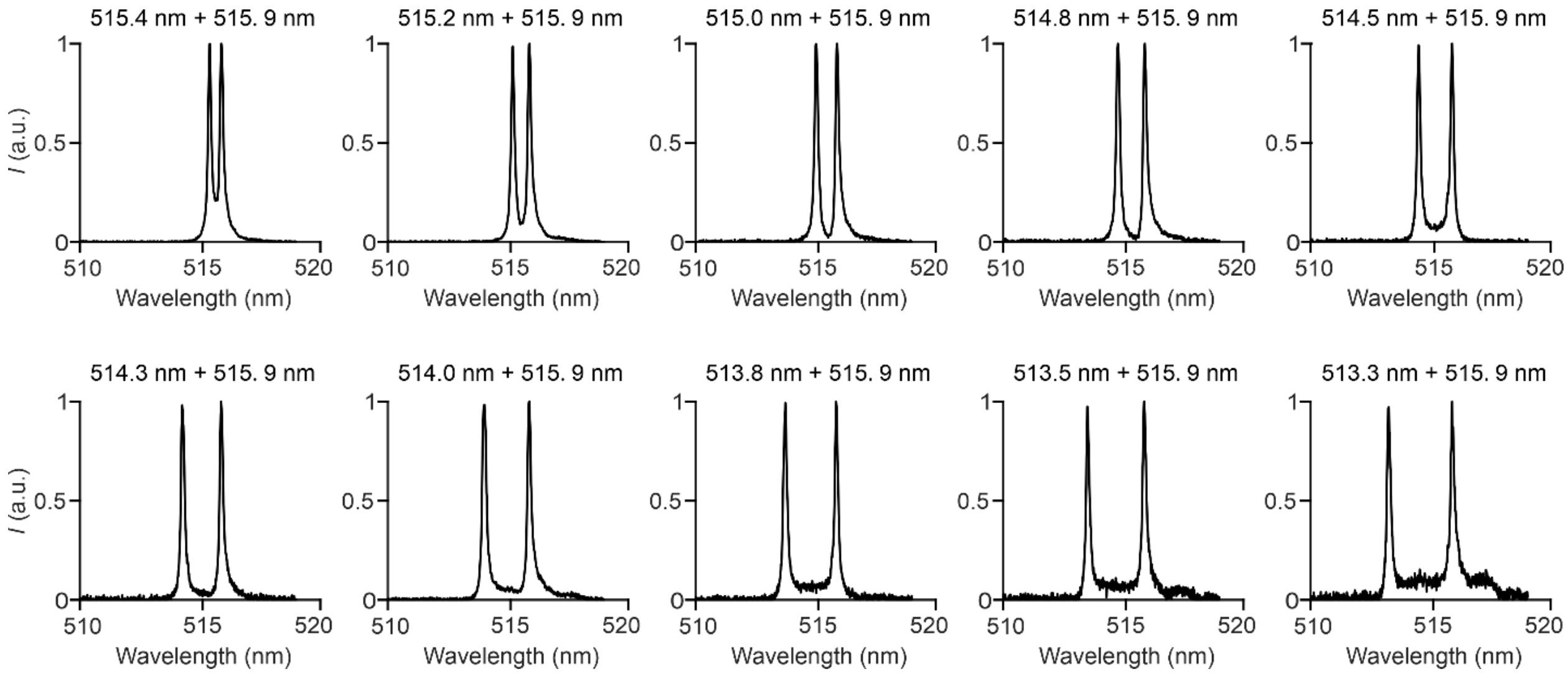

**Extended Data Fig. 2 | Two-color laser waves.** Measured laser spectra for each electron tomography data of Fig. 2a. The measured central wavelengths of the peaks are indicated above each spectrum. The full width at half maximum of all peaks is < 0.2 nm.

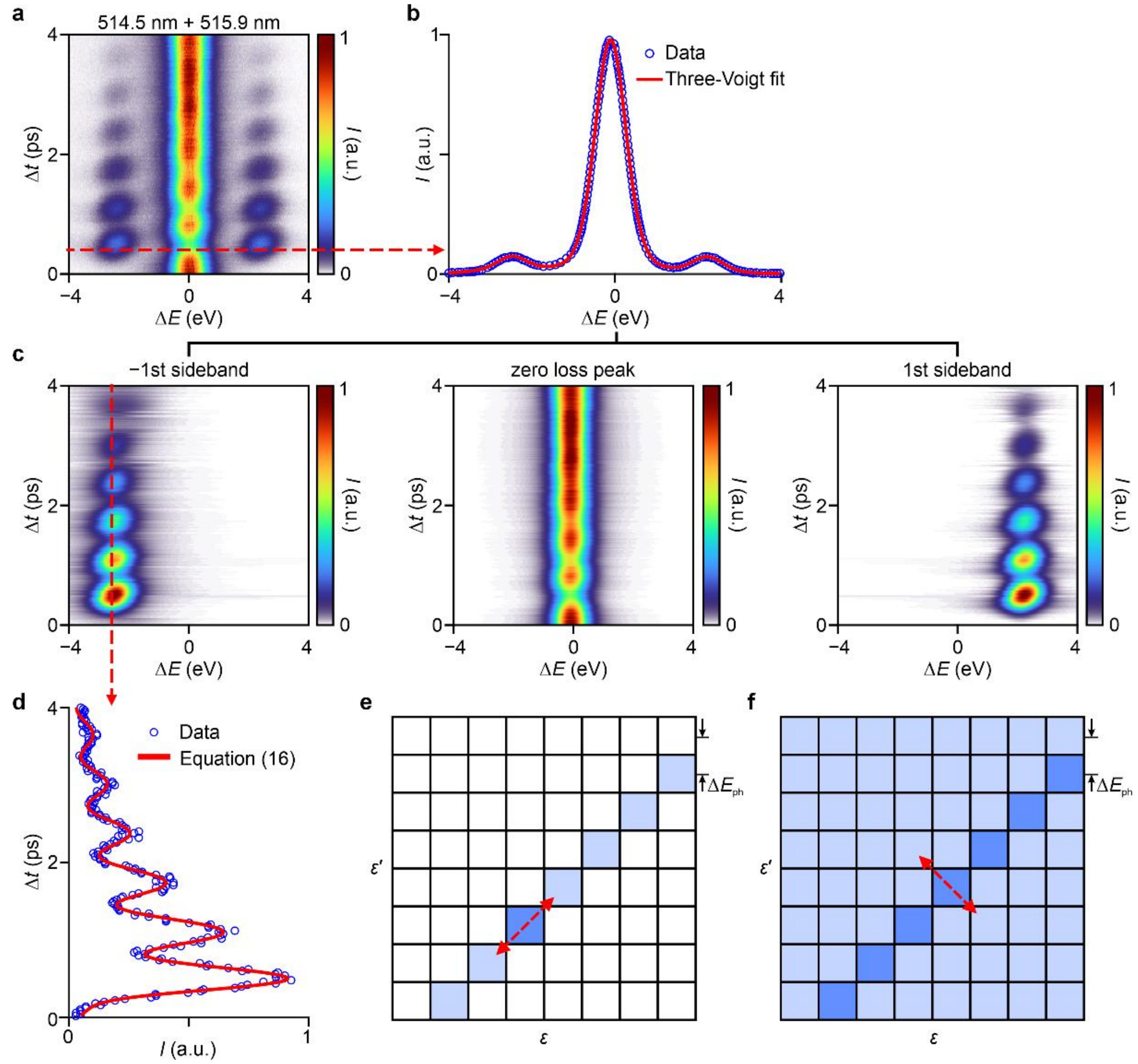

**Extended Data Fig. 3 | Workflow of the quantum state tomography.** **a**, Experimental data. **b**, Fit of the measured data (blue circles) with a sum of three Voigt functions (red curve) for each time delay (red dashed arrow). **c**, Resulting time-dependent interferograms of the -1th sideband (left), zero-loss peak (middle) and +1st sideband (right). **d**, Fit (red) of the time-delay oscillation (blue circles) at a fixed $\Delta E$ (red dashed arrow) using Eq. (16). **e**, Density matrix element extracted from this fit (dark blue square). Further matrix elements (light blue squares) are obtained by repeating the above procedure for other $\Delta E$. **f**, Further density matrix elements (light blue) are measured by tuning $\Delta E_{\mathrm{ph}}$ of the probe laser (light blue squares). For details, see Materials and Methods.

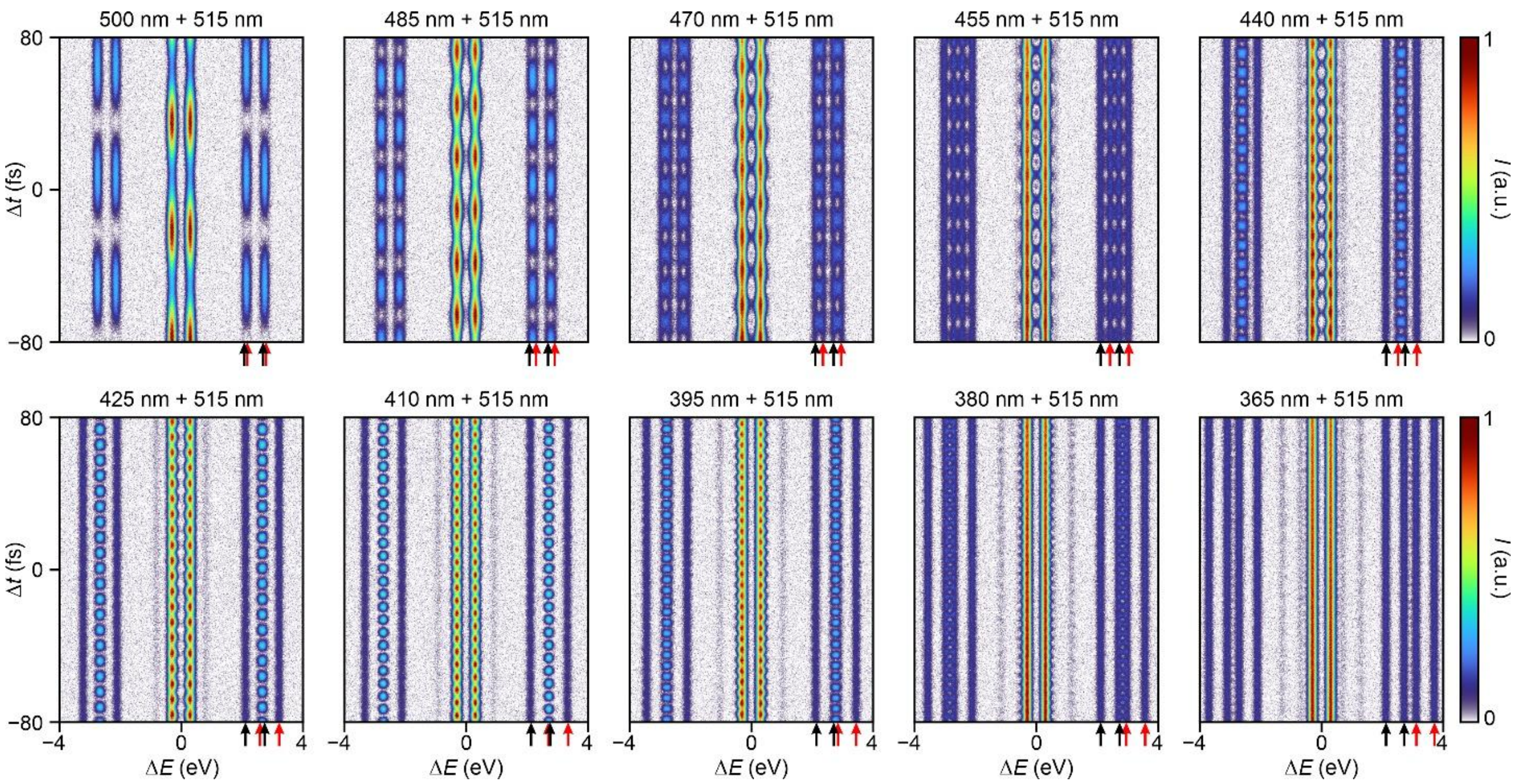

**Extended Data Fig. 4 | Simulation of the expected experimental observations for a free-electron cat state.** We assume a free-electron cat state in the form of $|\psi\rangle = |E_{el}+\Delta E\rangle + |E_{el}-\Delta E\rangle$, where $|E_{el}\rangle$ is a pure state at a central kinetic energy of $E_{el}$ = 180 keV; $\Delta E$ = 0.3 eV is the energy separation between the two partial states. The different panels show simulated spectral shearing interferometry data for varying laser wavelengths. The first order sideband around 3.5 eV is used to retrieve the quantum state (see Fig. 3). Black and red arrows denote the positions of the first sidebands that are created from the two detuned laser wavelengths. For a small detuning (500 nm and 515 nm, for example), we probe the coherence within $|E_{el}+\Delta E\rangle$ and $|E_{el}-\Delta E\rangle$, that is, the quantum properties within each part of the cat state. For a larger detuning (425 nm and 515 nm, for example), we probe the coherence between $|E_{el}+\Delta E\rangle$ and $|E_{el}-\Delta E\rangle$, that is, the quantum properties of the cate state.

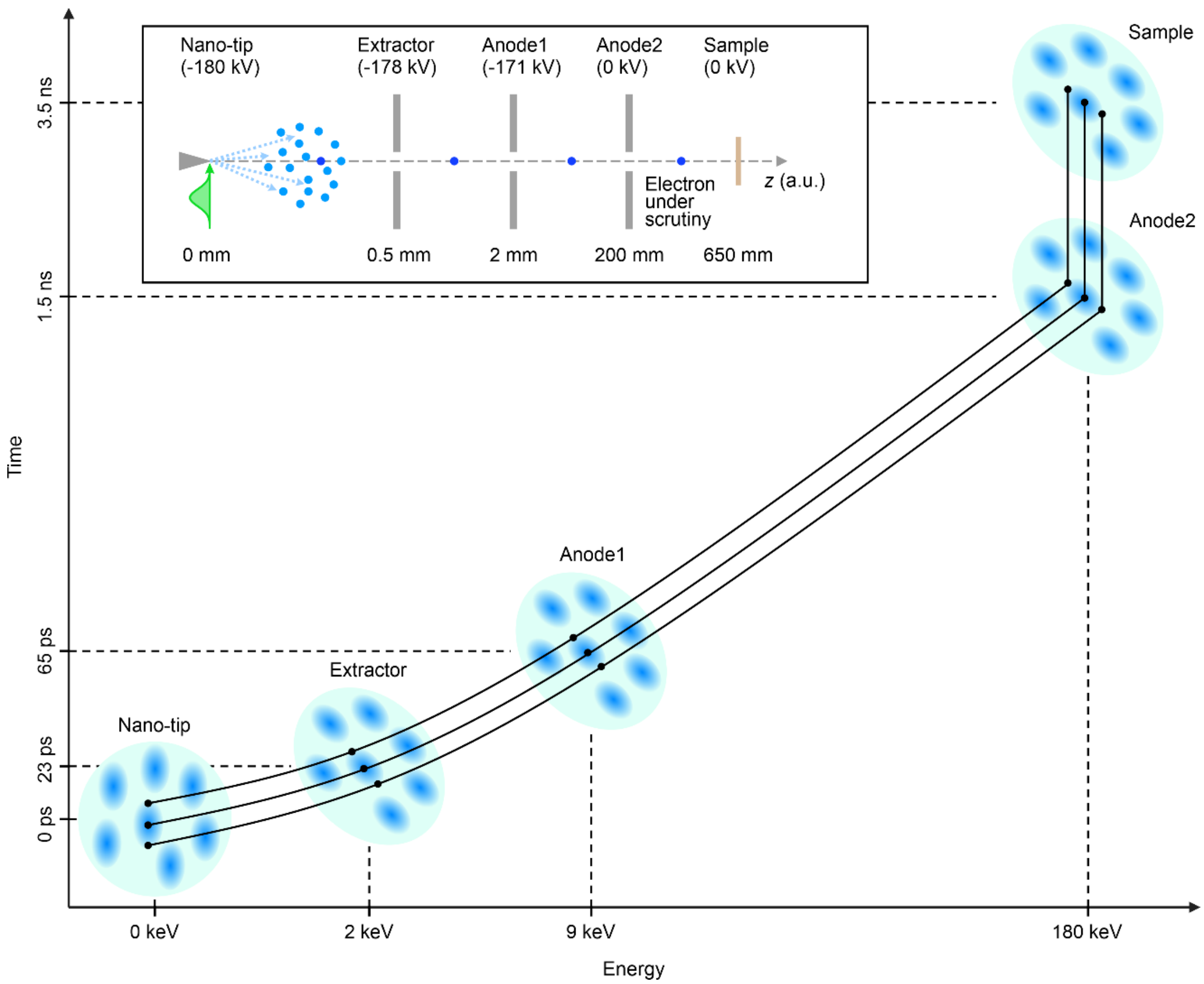

**Extended Data Fig. 5 | Schematic of our semiclassical model.** Light blue ellipses denote phase-space distributions of electron ensembles and dark blue ellipses denote pure states of individual electrons, as a function of electron energy and time. Vertical and horizontal dashed lines mark the central energies and times of the electrons at the emitter tip, the extractor, anode 1 and anode 2. The electrostatic potentials are -180 kV, -178 kV, -171 kV, 0 kV and 0 kV for the tip, extractor, anode 1, anode 2 and sample position, respectively, and their coordinates are 0 mm, 0.5 mm, 2 mm, 200 mm and 650 mm. Black dots mark point-like parts of an electron pure state and black solid curves illustrate the paths of these point-like parts in the time-energy domain. Inset: Multiple free electrons (light blue dots and dark blue dots) are photoemitted from the emitter tip (grey triangle) by each ultrafast laser pulse (green). Among these electrons, only one electron under scrutiny (dark blue) passes through the following apertures (grey rectangles) and arrives at the sample position (brown) for diagnostics with spectral shearing interferometry.

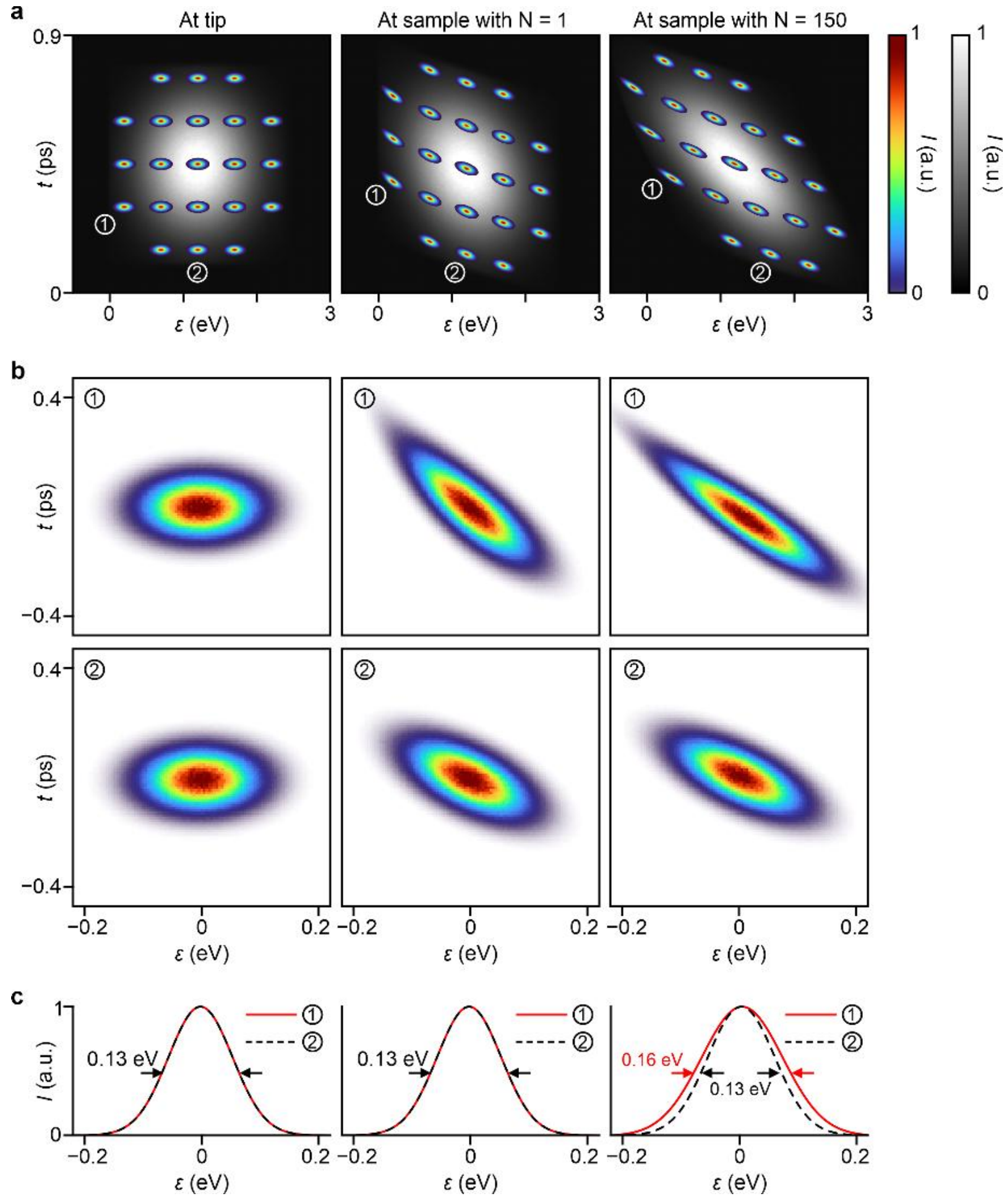

**Extended Data Fig. 6 | Phase space distributions and coherence lengths in the regime of space charge. a**, Mixed plots of the phase space distributions of the incoherent electron ensembles (grey) and pure example wavepackets (colored). Left column, electrons at the tip; middle column, electrons detected at the electron microscope's sample stage in the absence of the Coulomb interactions; right column, electrons at the sample stage for 150 electrons in the electron cloud and including Coulomb interactions. **b**, Zoom-in plots for pure states ① and ② as shown in a. **c**, Spectra of the pure wavepackets, obtained by integrating phase space along the $t$ axis. Left panel: The energy bandwidth is 0.13 eV (FWHM) for electron wavepackets at the tip, corresponding to a temporal coherence length of 14 fs (FWHM). Middle panel: Dispersive propagation in the microscope introduces chirp to the pure wavepackets, but their energy bandwidths remain almost unchanged. Right panel: With many-body Coulomb effects, the pure energy bandwidth of electron wavepacket ① is slightly broadened to 0.16 eV (FWHM), corresponding to a slightly reduced temporal coherence length of 11 fs (FWHM), while the energy width of wavepacket ② remains mostly unchanged. We hence clearly see that electrons at different locations within the space charge cloud obtain different decoherence properties.